\documentclass[11pt,ams]{article}
\usepackage{geometry}
\usepackage{dsfont}
\usepackage{amsmath}
\usepackage{amsfonts}
\usepackage{amssymb}
\usepackage{graphicx}

\begin{document}

\date{}
\title{\textbf{Renormalizability of the linearly broken formulation of the BRST symmetry in presence of the Gribov horizon in Landau gauge Euclidean Yang-Mills theories}}
\author{\textbf{M.~A.~L.~Capri}$^{a}$\thanks{%
capri@ufrrj.br}\,\,, \textbf{A.~J.~G\'{o}mez}$^{b}$\thanks{%
ajgomez@uerj.br}\,\,, \textbf{M.~S.~Guimaraes}$^{b}$\thanks{%
msguimaraes@uerj.br}\,\,, \textbf{V.~E.~R.~Lemes}$^{b}$\thanks{%
vitor@dft.if.uerj.br}\,\,, \\
%EndAName
\textbf{S.~P.~Sorella}$^{b}$\thanks{%
sorella@uerj.br}\, \thanks{%
Work supported by FAPERJ, Funda{\c{c}}{\~{a}}o de Amparo {\`{a}} Pesquisa do
Estado do Rio de Janeiro, under the program \textit{Cientista do Nosso Estado%
}, E-26/101.578/2010.}\,\,,\,\,\textbf{D.~G.~Tedesco}$^{b}$\thanks{%
dgtedesco@uerj.br}\,\,\\
[2mm] \textit{{\small $^{a}$ UFRRJ $-$ Universidade Federal Rural do Rio de
Janeiro}}\\
\textit{{\small Departamento de F\'{\i}sica $-$ Grupo de F\'{\i}sica Te\'{o}%
rica e Matem\'{a}tica F\'{\i}sica}}\\
\textit{\small BR 465-07, 23890-971, Serop\'edica, RJ, Brasil.}\\
\textit{{\small {$^{b}$ UERJ $-$ Universidade do Estado do Rio de Janeiro}}}%
\\
\textit{{\small {Instituto de F\'{\i}sica $-$ Departamento de F\'{\i}sica Te%
\'{o}rica}}}\\
\textit{{\small {Rua S{\~a}o Francisco Xavier 524, 20550-013 Maracan{\~a},
Rio de Janeiro, RJ, Brasil.}}}}
\maketitle

\begin{abstract}
\noindent In previous work  \cite{Capri:2010hb} we have shown that the soft breaking of the BRST symmetry arising within the Gribov-Zwanziger framework can be converted into a linear breaking, while preserving the nilpotency of the BRST operator. Due to its compatibility with the Quantum Action Principle, the linearly broken BRST symmetry directly translates into a set of Slavnov-Taylor identities. We show that these identities guarantee the multiplicative renormalizability of both Gribov-Zwanziger and Refined Gribov-Zwanziger theories to all orders. The known property that only two renormalization factors are needed  is recovered. The non-renormalization theorem of the gluon-ghost-antighost vertex as well as the renormalization factor of the Gribov parameter are  derived within the linearly broken formulation.
\end{abstract}

\baselineskip=13pt

\newpage

\section{Introduction}

The issue of the BRST symmetry versus the color confinement  is one of the highlight of the current debate on the infrared behavior of Yang-Mills theories. \\\\Needless to say, the BRST symmetry is a fundamental tool in order to: $i)$ prove the renormalizability of the theory, $ii)$ identify the physical  subspace whose states have positive norm, $iii)$ guarantee that the restriction to the physical subspace of the scattering operator $S$ is unitary. All this applies to the perturbative regime and to nonconfining gauge theories for which the asymptotic fields and their corresponding elementary particles can be safely introduced. \\\\Things become much more complicated when the theory under investigation is a confining theory, such as pure, {\it i.e.} quarkless, Yang-Mills theories. Gluons are not part of the spectrum, due to the nonperturbative phenomenon of color confinement. The contact with the physical spectrum of the theory is encoded in the  correlation functions of suitable composite colorless operators built out from the elementary gluon field. At present, it is safe to state that we lack a clear understanding of the role played by the BRST symmetry in a confining theory. \\\\ In this work we aim at  pursuing the investigation on the BRST symmetry within the so called Gribov-Zwanziger framework \cite{Gribov:1977wm,Zwanziger:1988jt,Zwanziger:1989mf}, which enables us to take into account the nonperturbative effect of the Gribov copies by restricting the domain of integration in the functional integral to the first Gribov horizon, see \cite{Dudal:2009bf} for a review.   Remarkably, this restriction can be implemented within an Euclidean field theory framework. The corresponding action, known as the Gribov-Zwanziger action, enjoys the property of being local and renormalizable \cite{Zwanziger:1988jt,Zwanziger:1989mf,Maggiore:1993wq,Dudal:2005na,Gracey:2006dr,Dudal:2010fq}. As a consequence of the restriction of the domain of integration to the Gribov horizon, the gluon two-point correlation function gets deeply modified in the infrared region, namely
\begin{equation}
\left\langle A_{\mu }^{a}(k)A_{\nu }^{b}(-k)\right\rangle =\delta ^{ab} \left( \delta _{\mu \nu }-\frac{k_{\mu }k_{\nu }}{
k^{2}}\right)
\frac{
k^{2}}{k^{4}+\gamma ^{4}} = \frac{\delta^{ab}}{2}  \left( \delta _{\mu \nu }-\frac{k_{\mu }k_{\nu }}{
k^{2}}\right) \left( \frac{1}{k^2+i\gamma^2} + \frac{1}{k^2-i\gamma^2} \right)  \;, \label{gribprop}
\end{equation}
where $\gamma$ is the Gribov parameter \cite{Gribov:1977wm,Zwanziger:1988jt,Zwanziger:1989mf,Dudal:2009bf}. Expression \eqref{gribprop} displays complex pole, so that it cannot describe a physical particle, being suitable for a confining phase.\\\\The issue of the BRST symmetry within the Gribov-Zwanziger framework has been much studied in recent years \cite{Baulieu:2009ha,Dudal:2008sp,Baulieu:2008fy,Sorella:2009vt,Dudal:2009xh,Kondo:2009qz,Dudal:2010hj, Zwanziger:2010iz,Capri:2010hb}, as summarized below. The first observation which can be made is that the Gribov-Zwanziger action does not enjoy exact BRST invariance. The usual BRST symmetry of the Faddeev-Popov action turns out to be broken by terms proportional to the Gribov parameter $\gamma^2$. Moreover, as the resulting breaking term is of dimension two in the fields, it is a soft breaking \cite{Dudal:2008sp,Baulieu:2008fy}. As such, it can be kept under control at the quantum level. This amounts to treat the breaking as a composite operator which is introduced into the theory through a set of external fields. This enables us to write down generalized Slavnov-Taylor identities which guarantee the renormalizability of the Gribov-Zwanziger theory to all orders. These generalized Slavnov-Taylor applies as well to the renormalization of gauge invariant composite operators \cite{Dudal:2009zh}. In \cite{Sorella:2009vt,Kondo:2009qz}, the softly broken BRST symmetry of the Gribov-Zwanziger action has been cast into an exact invariance which is, however, non-local. Further, in \cite{Dudal:2010hj} this non-local invariance has been reformulated as an exact local symmetry by introducing a suitable set of additional localizing fields. Though, the resulting BRST operator is not nilpotent. Attempts at interpreting the breaking of the BRST symmetry as a spontaneous symmetry breaking can be found in \cite{Zwanziger:2010iz}. \\\\Recently, we have been able to show that the soft breaking of the BRST symmetry exhibited by the Gribov-Zwanziger theory can be converted into a linear breaking \cite{Capri:2010hb}, namely the Gribov-Zwanziger action admits an equivalent formulation for which the BRST symmetry is only linearly broken. This is a useful observation, due to the fact that a linear breaking is compatible with the Quantum Action Principle\footnote{For an introduction to the Quantum Action Principle, see \cite{Piguet:1995er} and references therein.}, while preserving the nilpotency of the BRST operator. As such, the linearly broken BRST symmetry can be directly translated into a system of Slavnov-Taylor identities. Furthermore,  similarly to the case of the dimension two BRST soft breaking  \cite{Dudal:2010fq}, the linearly broken formulation allows us tow write down additional functional identities, so that a large set of Ward identities can be established. \\\\Here, we prove that these Ward identities ensure the multiplicative renormalizability of both the Gribov-Zwanziger and Refined Gribov-Zwanziger \cite{Dudal:2007cw,Dudal:2008sp} actions  within the linearly broken BRST formulation. As we shall see, only two renormalization factors are needed, which will be identified with the renormalization factor of the gauge coupling constant $g$, $Z_g$, and of the gauge field $A_\mu^a$, $Z_A$. We point out that the linearly broken formulation enables us to establish a particular Ward identity which governs the dependence of the allowed counterterms from the Gribov parameter $\gamma^2$. This Ward identity provides a simple understanding of the nonrenormalization properties of the Gribov parameter, expressing the fact that the renormalization factor $Z_{\gamma^2}$ is not an independent quantity, according to the relation $Z_{\gamma^2}= Z_g^{-1/2} Z_A^{-1/4}$. \\\\The work is organized as follows. In Sect.2 a brief review of the Gribov-Zwanziger action within the linear BRST broken formulation is provided. In Sect.3. the set of Ward identities is established. Sect. 4 deals with the algebraic characterization of the allowed invariant counterterm. We show that it can  be reabsorbed into the starting action by a multiplicative redefinition of the parameters and fields, thus establishing the renormalizability of the Gribov-Zwanziger action. In Sect.5 the renormalizability of the Refined Gribov-Zwanziger action is addressed.  In Sect.6 we outline a few observations on the breaking of the BRST symmetry in a confining theory, within the context of the Refined Gribov-Zwanziger framework.

\section{Linearly  broken BRST formulation of the Gribov-Zwanziger theory}

In this section we give a short account of the BRST linearly broken formulation of the Gribov-Zwanziger action.  Let us start with the conventional formulation of the theory, which exhibits a soft BRST breaking.

\subsection{The Gribov-Zwanziger action and its soft BRST breaking}

As already mentioned, the Gribov-Zwanziger action implements in a local way the restriction of the domain of integration in the functional integral to the Gribov region $\Omega$, which is the set of all transverse gauge configurations for which the Faddeev-Popov operator, $-\partial_\mu \left(\partial _{\mu }\delta
^{ab}+gf^{acb}A_{\mu }^{c}\right) $, is strictly positive
\begin{equation}
\Omega = \{ \;A^a_\mu\;, \;\;\partial_\mu A^a_\mu=0\;, \;\; -\partial_\mu \left(\partial _{\mu }\delta
^{ab}+gf^{acb}A_{\mu }^{c}\right) >0 \;\}  \;. \label{omega}
\end{equation}
The boundary, $\partial \Omega$, of the region $\Omega$, where the first vanishing eigenvalue of the Faddeev-Popov operator shows up, is called the Gribov horizon.
The action of the theory is given by the following local expression
\begin{eqnarray}
S_{GZ} & = & \int d^4 x \left( \frac1{4} F_{\mu\nu}^{a}F_{\mu\nu}^{a} +
ib^a\partial_{\mu}A^a_\mu + {\bar c}^a \partial_\mu D_{\mu}^{ab} c^b \right)
\notag \\
{\ }{\ }{\ } & + &\int d^4x \left( - {\bar \varphi}^{ac}_\mu \partial_\nu
D_{\nu}^{ab} \varphi^{bc}_{\mu} + {\bar \omega}^{ac}_\mu \partial_\nu
D_{\nu}^{ab} \omega^{bc}_{\mu} + g f^{amb} (\partial_{\nu}{\bar \omega}%
^{ac}_{\mu}) (D^{mp}_{\nu} c^p)\varphi^{bc}_{\mu} \right)  \notag \\
{\ }{\ }{\ } & +& \int
d^4x\left(\gamma^2\,g\,f^{abc}A_\mu^{a}(\varphi_\mu^{bc}-\bar{\varphi}%
_\mu^{bc})-d(N^2-1)\gamma^4 \right)  \notag \\
& = &\frac1{4}\int d^4 x F_{\mu\nu}^{a}F_{\mu\nu}^{a} + s\int d^4 x \left(%
\bar{c}^a\partial_{\mu} A^a_{\mu} - \bar{\omega}_\mu^{ac}\partial_\nu
D^{ab}_\nu \varphi_{\mu}^{bc}\right) + S_{\gamma} \;,  \label{GZact}
\end{eqnarray}
with
\begin{align}
S_{\gamma}&=\int d^4x\left(\gamma^2\,g\,f^{abc}A_\mu^{a}(\varphi_\mu^{bc}-%
\bar{\varphi}_\mu^{bc})-d(N^2-1)\gamma^4 \right) \;,  \label{BRSbreak}
\end{align}
where $N$ is the number of colors and $d=4$ is the spacetime dimension. The fields $({\bar \varphi}^{ab}_\mu,  \varphi^{ab}_\mu )$ are a pair of complex conjugate commuting fields, while $({\bar \omega}^{ab}_\mu,  \omega^{ab}_\mu )$ are anticommuting fields. Each of these field has $4(N^2-1)^2$ independent components. The
Gribov parameter $\gamma ^{2}$ is not an independent parameter of the theory. It is determined in a self-consistent way by the gap
equation \cite{Zwanziger:1988jt,Zwanziger:1989mf,Dudal:2009bf}
\begin{equation}
\frac{\partial \mathcal{E}_{vac}}{\partial \gamma ^{2}}=0\;,
\end{equation}
where $\mathcal{E}_{vac}$ is the vacuum energy
\begin{equation}
e^{-\mathcal{E}_{vac}}=\int \left[ d\phi \right] e^{-S_{GZ}}\;,
\end{equation}
and $\left[ d\phi \right] $ stands for integration over all fields entering the expression \eqref{GZact}.
In the absence of the term $S_{\gamma }$, which would imply the removal of the Gribov horizon,  the action \eqref{GZact} turns out to be equivalent to the ordinary Faddeev-Popov action \cite{Dudal:2009bf}. As such, it displays exact BRST
symmetry
\begin{align}
sA_{\mu }^{a}& =-D_{\mu }^{ab}c^{b}=-(\partial _{\mu }\delta
^{ab}+gf^{acb}A_{\mu }^{c})c^{b}\;,  \notag \\
sc^{a}& =\frac{g}{2}f^{acb}c^{b}c^{c}\;,  \notag \\
s\bar{c}^{a}& =ib^{a}\;,\qquad sb^{a}=0\;,  \notag \\
s\bar{\omega}_{\mu }^{ab}& =\bar{\varphi}_{\mu }^{ab}\;,\qquad s\bar{\varphi}%
_{\mu }^{ab}=0\;,  \notag \\
s\varphi _{\mu }^{ab}& =\omega _{\mu }^{ab}\;,\qquad s\omega _{\mu
}^{ab}=0\;,  \label{BRS}
\end{align}%
with
\begin{equation}
s\int d^{4}x\left( \frac{1}{4}F_{\mu \nu }^{a}F_{\mu \nu }^{a}+s\left( \bar{c%
}^{a}\partial _{\mu }A_{\mu }^{a}-\bar{\omega}_{\mu }^{ac}\partial _{\nu
}D_{\nu }^{ab}\varphi _{\mu }^{bc}\right) \right) =0\;.
\end{equation}
However, in the presence of the horizon, thus $\gamma^2\neq 0$, the Gribov-Zwanziger action is not left invariant by the BRST
transformations, eqs.\eqref{BRS}, which are broken by the term $S_{\gamma }$%, namely
\begin{equation}
sS_{GZ}=sS_{\gamma }=\gamma ^{2}\,\int d^{4}x\left( -g\,f^{abc}(D_{\mu
}^{ad}c^{d})(\varphi _{\mu }^{bc}-\bar{\varphi}_{\mu
}^{bc})+g\,f^{abc}A_{\mu }^{a}\omega _{\mu }^{bc}\right) \;.
\label{BRSactbreak}
\end{equation}
Notice that, being of dimension two in the fields, the breaking term is soft.

%%%%%%%%%%%%%%%%%%%%%%%%%%%%%%%%%%%%%%%%%%%%%%%%%%%%%%%%%%%%%

\subsection{Converting the soft breaking into a linear breaking}

%%%%%%%%%%%%%%%%%%%%%%%%%%%%%%%%%%%%%%%%%%%%%%%%%%%%%%%%%%%%%
As it has been pointed out in  \cite{Capri:2010hb}, the soft breaking of the BRST symmetry exhibited by the Gribov-Zwanziger action can be converted into a linear breaking. To that purpose, we have introduced two sets of BRST quartets of auxiliary fields,
$(\bar{\mathcal{C}}_{\mu \nu }^{ab},\lambda _{\mu \nu }^{ab},\eta
_{\mu \nu }^{ab},\mathcal{C}_{\mu \nu }^{ab})$ and $(\bar{\rho}_{\mu \nu
}^{ab},\bar{\lambda}_{\mu \nu }^{ab},\bar{\eta}_{\mu \nu }^{ab},\rho _{\mu
\nu }^{ab})$,  transforming as
\begin{align}
s\bar{\mathcal{C}}_{\mu \nu }^{ab}& =\lambda _{\mu \nu }^{ab}\;,\qquad
s\lambda _{\mu \nu }^{ab}=0\;,\qquad s\eta _{\mu \nu }^{ab}=\mathcal{C}_{\mu
\nu }^{ab}\;,\qquad s\mathcal{C}_{\mu \nu }^{ab}=0\;,  \notag \\
s\bar{\rho}_{\mu \nu }^{ab}& =\bar{\lambda}_{\mu \nu }^{ab}\;,\qquad s\bar{%
\lambda}_{\mu \nu }^{ab}=0\;,\qquad s\bar{\eta}_{\mu \nu }^{ab}=\rho _{\mu
\nu}^{ab}\;,\qquad s\rho _{\mu \nu }^{ab}=0\;.  \label{BRS2}
\end{align}
Here, $(\lambda _{\mu \nu }^{ab},\eta _{\mu \nu }^{ab},\bar{%
\lambda}_{\mu \nu }^{ab},\bar{\eta}_{\mu \nu }^{ab})$ are commuting fields,
while $(\bar{\mathcal{C}}_{\mu \nu }^{ab},\mathcal{C}_{\mu \nu }^{ab},
\bar{\rho}_{\mu \nu }^{ab},\rho _{\mu \nu }^{ab})$ are anticommuting ones.\\\\
Thus, as discussed in
details in \cite{Capri:2010hb}, the Gribov-Zwanziger action \eqref{GZact} can be cast into  the equivalent form \begin{align}
S_{GZ}^{lin}& =\frac{1}{4}\int d^{4}x\,F_{\mu \nu }^{a}F_{\mu \nu
}^{a}+s\int d^{4}x\left( \bar{c}^{a}\partial _{\mu }A_{\mu }^{a}-\bar{\omega}%
_{\mu }^{ac}\partial _{\nu }D_{\nu }^{ab}\varphi _{\mu }^{bc}\right)  \notag
\\
& +s\int d^{4}x\left( -\bar{\mathcal{C}}_{\mu \nu }^{cd}D_{\mu }^{cb}\varphi
_{\nu }^{bd}-\bar{\mathcal{C}}_{\mu \nu }^{ab}\eta _{\mu \nu }^{ab}-\bar{%
\mathcal{C}}_{\mu \nu }^{ab}\bar{\eta}_{\mu \nu }^{ab}+\bar{\eta}_{\mu \nu
}^{cd}D_{\mu }^{cb}\bar{\omega}_{\nu }^{bd}-\bar{\rho}_{\mu \nu }^{ab}\bar{%
\eta}_{\mu \nu }^{ab}\right)  \notag \\
& +\int d^{4}x\left( \gamma ^{2}\eta _{\mu \nu }^{ab}\delta ^{ab}\delta
_{\mu \nu }+\gamma ^{2}\bar{\lambda}_{\mu \nu }^{ab}\delta ^{ab}\delta _{\mu
\nu }\right) \;.  \label{GZactlinear}
\end{align}
To prove the equivalence between the two formulations, \eqref{GZact} and \eqref{GZactlinear}, one proceeds \cite{Capri:2010hb} by integrating  out the fields $(\lambda _{\mu
\nu }^{ab},\eta _{\mu \nu }^{ab})$ and $(\bar{\lambda}_{\mu \nu }^{ab},\bar{%
\eta}_{\mu \nu }^{ab})$, a task easily done due to the way in which these fields enter the action \eqref{GZactlinear}. Further,  the fields $(\bar{\mathcal{C}}_{\mu \nu }^{ab},\mathcal{C}_{\mu
\nu }^{ab})$, $(\bar{\rho}_{\mu \nu }^{ab},\rho _{\mu \nu }^{ab})$ can be completely decoupled from the theory
by making a suitable field redefinition \cite{Capri:2010hb}. This leads to the equivalence between the two formulations  eq.\eqref{GZact}, and eq.\eqref{GZactlinear} .\\\\
By making use of the BRST transformations (\ref{BRS}) and (\ref{BRS2}), it is apparent that, instead of a soft breaking,   the action \eqref{GZactlinear} exhibits a linear breaking of the BRST symmetry, \textit{i.e.} the resulting breaking term is linear in the fields, namely
\begin{equation}
sS_{GZ}^{lin}=\gamma ^{2}\,\int d^{4}x\,\delta ^{ab}\delta _{\mu \nu }%
\mathcal{C}_{\mu \nu }^{ab}\;. \label{linbs}
\end{equation}
Notice also that the BRST operator $s$, as defined by eqs.\eqref{BRS} and eqs.\eqref{BRS2}, is nilpotent
\begin{equation}
s^2 = 0 \;. \label{nilp}
\end{equation}
As a linear breaking is compatible with the Quantum Action Principle  \cite{Piguet:1995er},   it can be directly translated into Slavnov-Taylor identities. This is the task of the next section.
%%%%%%%%%%%%%%%%%%%%%%%%%%%%%
\section{Ward Identities}
%%%%%%%%%%%%%%%%%%%%%%%%%%%%%

In order to derive the Ward Identities obeyed by of the action \eqref{GZactlinear}, it turns out to be useful to introduce the multi-index notation  \cite{Zwanziger:1988jt,Zwanziger:1989mf,Maggiore:1993wq,Dudal:2005na,Dudal:2010fq}.
\begin{eqnarray}
\left(\varphi^{ab}_{\mu},\bar\varphi^{ab}_{\mu},\omega^{ab}_{\mu},\bar\omega^{ab}_{\mu}\right)
&\equiv&\left(\varphi^{a}_{i},\bar\varphi^{a}_{i},\omega^{a}_{i},\bar\omega^{a}_{i}\right)\;,\nonumber\\
\left(\bar{\mathcal{C}}_{\mu \nu }^{ab},\lambda _{\mu \nu }^{ab},\eta
_{\mu \nu }^{ab},\mathcal{C}_{\mu \nu }^{ab}\right)&\equiv&
\left(\bar{\mathcal{C}}_{\mu i }^{a},\lambda _{\mu i }^{a},\eta
_{\mu i }^{a},\mathcal{C}_{\mu i }^{a}\right)\;,\nonumber\\
\left(\bar{\rho}_{\mu \nu
}^{ab},\bar{\lambda}_{\mu \nu }^{ab},\bar{\eta}_{\mu \nu }^{ab},\rho _{\mu
\nu }^{ab}\right)&\equiv&
\left(\bar{\rho}_{\mu i
}^{a},\bar{\lambda}_{\mu i }^{a},\bar{\eta}_{\mu i}^{a},\rho _{\mu
i }^{a}\right)\;,
\end{eqnarray}
where $i\equiv\{a,\mu\}=1,\dots,f$, with $f=d(N^{2}-1)$ and $d=4$ being the dimension of the Euclidean spacetime. The explicit  expression of  the action $S_{GZ}^{lin}$ reads
\begin{eqnarray}
S_{GZ}^{lin} &=&\int d^{4}x\left( \frac{1}{4}F_{\mu \nu }^{a}F_{\mu \nu
}^{a}+ib^{a}\partial _{\mu }A_{\mu }^{a}+\bar{c}^{a}\partial _{\mu }D_{\mu
}^{ab}c^{b}-\bar{\varphi}_{i}^{a}\partial _{\mu }D_{\mu }^{ab}\varphi
_{i}^{b}+\bar{\omega}_{i}^{a}\partial _{\mu }D_{\mu }^{ab}\omega
_{i}^{b}\right.  \notag \\
&&\left. +gf^{amb}(\partial _{\nu }\bar{\omega}_{i}^{a})(D_{\nu
}^{mp}c^{p})\varphi _{i}^{b}+\lambda _{\mu i}^{a}D_{\mu }^{ab}\varphi
_{i}^{b}-\bar{\eta}_{\mu i}^{a}D_{\mu }^{ab}\bar{\varphi}_{i}^{b}+gf^{abc}%
\bar{\eta}_{\mu i}^{c}(D_{\mu }^{ap}c^{p})\bar{\omega}_{i}^{b}\right.  \notag
\\
&&\left. -\lambda _{\mu i}^{a}\bar{\eta}_{\mu i}^{c}-\mathcal{\bar{C}}_{\mu
i}^{a}D_{\mu }^{ab}\omega _{i}^{b}-\rho _{\mu i}^{a}D_{\mu }^{ab}\bar{\omega}%
_{i}^{b}+\mathcal{\bar{C}}_{\mu i}^{c}\left[ \mathcal{C}_{\mu i}^{c}+\rho
_{\mu i}^{c}+gf^{abc}(D_{\mu }^{ap}c^{p})\varphi _{i}^{b}\right] \right.
\notag \\
&&\left. -\rho _{\mu i}^{a}\bar{\rho}_{\mu i}^{a}+\bar{\lambda}_{\mu
i}^{a}(\gamma ^{2}\delta ^{ab}\delta _{\mu \nu }\delta _{\nu b}^{i}-\bar{\eta%
}_{\mu i}^{a})+\eta _{\mu i}^{a}(\gamma ^{2}\delta ^{ab}\delta _{\mu \nu
}\delta _{\nu b}^{i}-\lambda _{\mu i}^{a})\right)\;.  \label{action}
\end{eqnarray}
Following the set up of the Algebraic Renormalization  \cite{Piguet:1995er}, in order to account for the nonlinear BRST transformations of the gauge field $A^a_\mu$ and of the ghost field $c^a$, we introduce a pair of BRST external sources  $(\Omega^a_\mu, L^a)$ coupled to the composite operators  $(sA^a_\mu)$ and $(sc^a)$, namely
\begin{equation}
S_{\rm ext} = s \int d^4x\;\left( -\Omega^a_\mu A^a_\mu + L^a c^a \right) = \int d^{4}x\; \left( -\Omega _{\mu
}^{a}D_{\mu }^{ab}c^{b}+\frac{g}{2}f^{abc}L^{a}c^{b}c^{c}\right) \;. \label{extact}
\end{equation}
Thus, the complete action
\begin{equation}
\Sigma=S_{GZ}^{lin}+ S_{\rm ext}  \;.
\label{quant-GZlinear}
\end{equation}
fulfills the following set of Ward identities:
\begin{itemize}
\item{The linearly broken Slavnov-Taylor identity:
\begin{equation}
\mathcal{S}(\Sigma)=\gamma ^{2}\,\int d^{4}x\;\delta ^{ab}\delta
_{\mu \nu }\delta _{b\nu }^{i}\mathcal{C}_{\mu i}^{a}\;,  \label{st1}
\end{equation}
where
\begin{eqnarray}
\mathcal{S}(\Sigma) &=&\int d^{4}x\biggl(\frac{\delta \Sigma
}{\delta A_{\mu }^{a}}\frac{\delta\Sigma}{\delta
\Omega _{\mu }^{a}}+\frac{\delta \Sigma}{\delta L^{a}}\frac{
\delta \Sigma}{\delta c^{a}}+ib^{a}\frac{\delta \Sigma
}{\delta \bar{c}^{a}}
+\omega^{a}_{i}\frac{\delta\Sigma}{\delta\varphi^{a}_{i}}
+\bar{\varphi}^{a}_{i}\frac{\delta\Sigma}{\delta\bar\omega^{a}_{i}}\nonumber\\
&& {\ }{\ }{\ }{\ }{\ }+ \lambda _{\mu i}^{a}\frac{\delta \Sigma
}{\delta \bar{\mathcal{C}}_{\mu i}^{a}}+\mathcal{C}_{\mu i}^{a}
\frac{\delta \Sigma}{\delta \eta _{\mu i}^{a}}
+\bar{\lambda}_{\mu i}^{a}\frac{\delta \Sigma}{\delta \bar{\rho
}_{\mu i}^{a}}+\rho _{\mu i}^{a}\frac{\delta \Sigma}{\delta \bar{
\eta}_{\mu i}^{a}}\biggr)\,.\label{sti}
\end{eqnarray}
As one easily figures out, the presence of the linear breaking term in \eqref{st1} is inherited from the linearly broken BRST invariance, eq.\eqref{linbs}. Also, as a consequence of the nilpotency of the BRST operator $s$,  \eqref{nilp}, it follows that the linearized operator  $\mathcal{B}_{\Sigma}$ defined as
\begin{eqnarray}
\mathcal{B}_{\Sigma}&=&\int d^{4}x \biggl( \frac{
\delta\Sigma}{\delta A_{\mu}^a}\frac{\delta }{\delta
\Omega_{\mu}^a} + \frac{\delta\Sigma}{\delta \Omega_{\mu}^a}\frac{
\delta }{\delta A_{\mu}^a} + \frac{\delta\Sigma}{\delta L^a}\frac{
\delta }{\delta c^a} + \frac{\delta\Sigma}{\delta c^a}\frac{
\delta }{\delta L^a} + ib^a\frac{\delta }{\delta \bar{c}^a}  \notag \\
&&+\omega^{a}_{i}\frac{\delta}{\delta\varphi^{a}_{i}}
+\bar{\varphi}^{a}_{i}\frac{\delta}{\delta\bar\omega^{a}_{i}}
+\lambda_{\mu i}^a\frac{\delta}{\delta \bar{
\mathcal{C}}_{\mu i}^a}+ \mathcal{C}_{\mu i}^a\frac{\delta}{\delta \eta_{\mu
i}^a}+ \bar{\lambda}_{\mu i}^a\frac{\delta}{\delta \bar{\rho}_{\mu i}^a} +
\rho_{\mu i}^a\frac{\delta}{\delta \bar{\eta}_{\mu i}^a} \biggr)\;,
\label{linsti}
\end{eqnarray}
\begin{equation}
\mathcal{B}_{\Sigma}\mathcal{B}_{\Sigma}= 0 \;,
\label{nilp1}
\end{equation}
is nilpotent too. As we shall see later, this operator will enter the algebraic characterization of the most general invariant counterterm.}

\item{The equations of motion of the fields $b^a$, $\bar{c}^a$, $\mathcal{C}%
_{\mu i}^{a}$, $\bar\rho_{\mu i}^{a}$, $\eta_{\mu i}^{a}$ and $\bar{\lambda}%
_{\mu i}^{a}$:
\begin{equation}
\frac{\delta \Sigma}{\delta b^a} = i \partial_{\mu} A_{\mu}^a \;,
\qquad \frac{\delta \Sigma}{\delta \bar{c}^a} + \partial_{\mu}
\frac{\delta \Sigma}{\delta \Omega_{\mu}^a} = 0 \;,
\label{eqmotion1}
\end{equation}
\begin{equation}
\frac{\delta \Sigma}{\delta \eta_{\mu i}^{a}} = -\lambda_{\mu
i}^{a} + \gamma^2 \delta^{ab} \delta_{\mu\nu}\delta_{b\nu}^i \;, \qquad
\frac{\delta \Sigma}{\delta \bar{\lambda}_{\mu i}^{a}} = -\bar{%
\eta}_{\mu i}^{a} + \gamma^2 \delta^{ab} \delta_{\mu\nu}\delta_{b\nu}^i \;,
\label{eqmotion2}
\end{equation}
\begin{equation}
\frac{\delta \Sigma}{\delta \mathcal{C}_{\mu i}^{a}} = -\bar{%
\mathcal{C}}_{\mu i}^{a} \;, \qquad \frac{\delta \Sigma}{\delta
\bar\rho_{\mu i}^{a}} = {\rho}_{\mu i}^{a} \;.  \label{eqmotion3}
\end{equation}}

\item{The parametric Ward identity for the Gribov parameter $\gamma$:
\begin{align}
\frac{\partial \Sigma}{\partial \gamma^2} = \int
d^4x\;\delta^{ab} \delta_{\mu\nu}\delta_{b\nu}^i (\eta_{\mu i}^{a}+\bar{%
\lambda}_{\mu i}^{a}) \;.  \label{nonrenorm}
\end{align}
Notice that the right hand side of eq.\eqref{nonrenorm} is linear in the fields. Again, this term is a linear breaking, not affected by the quantum corrections. As already mentioned, this Ward identity has a special meaning. It implies that the allowed invariant counterterm does not depend explicitly on the parameter $\gamma^2$. As such, the  Ward identity \eqref{nonrenorm} provides a very simple understanding of the nonrenormalization properties of the Gribov parameter $\gamma^2$, as encoded in the relationship $Z_{\gamma^2}= Z_g^{-1/2} Z_A^{-1/4}$. We also point out that the possibility of writing down the Ward identity \eqref{nonrenorm} is a byproduct of the linear breaking formulation of the Gribov-Zwanziger action. }

\item{The local equation of motion of $\bar{\varphi}_i^a$:
\begin{align}
\Xi_{\bar\varphi}^{ai}(\Sigma)\equiv\frac{\delta \Sigma}{\delta \bar{\varphi}_i^a} +\partial_{\mu}%
\frac{\delta \Sigma}{\delta \lambda_{\mu i}^a} +gf^{abc}A_{\mu}^b%
\frac{\delta \Sigma}{\delta \bar{\lambda}_{\mu i}^c} =
\Delta^{ai}_{\bar{\varphi}} \;,  \label{eqmotion-barphi}
\end{align}
where $\Delta^{ai}_{\bar{\varphi}}$ is a  linear breaking, given by
\begin{align}
\Delta^{ai}_{\bar{\varphi}} = -\partial^2 \varphi_i^a -
\partial_{\nu}\eta_{\nu i}^a - \partial_{\mu}\bar{\eta}_{\mu i}^a -
g\gamma^2 f^{abc}A_{\mu}^c\delta_{b\mu}^i\;.  \label{delta-barphi}
\end{align}}

\item{The local equation of motion of $\bar{\omega}_i^a$:
\begin{align}
\Xi_{\bar\omega}^{ai}(\Sigma)\equiv\frac{\delta \Sigma}{\delta \bar{\omega}_i^a} +\partial_{\mu}
\frac{\delta \Sigma}{\delta \bar{\mathcal{C}}_{\mu i}^a}
+gf^{abc}A^{b}_{\mu}\frac{\delta\Sigma}{\delta\bar\rho^{c}_{\mu i}
} + gf^{abc}\left(\frac{\delta\Sigma}{\delta\bar\lambda^{b}_{\mu
i}}-\gamma^{2}\delta^{i}_{\mu b}\right) \frac{\delta\Sigma}{
\delta\Omega^{c}_{\mu}} = \Delta^{ai}_{\bar{\omega}} \;,
\label{eqmotion-baromega}
\end{align}
where
\begin{align}
\Delta^{ai}_{\bar{\omega}} = \partial^2 \omega_i^a + \partial_{\mu}\mathcal{C%
}_{\mu i}^a +\partial_{\mu}\rho^{a}_{\mu i} \;.  \label{delta-baromega}
\end{align}}

\item{The local equation of motion of $\varphi _{i}^{a}$:
\begin{eqnarray}
\Xi_{\varphi}^{ai}(\Sigma)&\equiv&\frac{\delta \Sigma}{\delta \varphi _{i}^{a}}-\partial _{\mu }
\frac{\delta \Sigma}{\delta \bar{\eta}_{\mu i}^{a}}-igf^{abc}
\bar{\varphi}_{i}^{b}\frac{\delta \Sigma}{\delta b^{c}}+gf^{abc}
\bar{\omega}_{i}^{b}\frac{\delta \Sigma}{\delta \bar{c}^{c}}
-gf^{abc}A_{\mu }^{b}\frac{\delta \Sigma}{\delta \eta _{\mu
i}^{c}}\nonumber\\
&&-gf^{acm}\frac{\delta \Sigma}{\delta \mathcal{C}_{\mu
i}^{c}}\frac{\delta \Sigma}{\delta \Omega _{\mu }^{m}}=\Delta
_{\varphi }^{ai}\;,\label{eqmotion-phi}
\end{eqnarray}
with
\begin{equation}
\Delta _{\varphi }^{ai}=-\partial ^{2}\bar{\varphi}_{i}^{a}+\partial _{\mu
}\lambda _{\mu i}^{a}+\partial _{\mu }\bar{\lambda}_{\mu i}^{a}-\gamma
^{2}gf^{abc}A_{\mu }^{b}\delta _{c\mu }^{i}\;. \label{delta-varphi}
\end{equation}}

\item{The local equation of motion of $\omega_i^a$:
\begin{align}
\Xi_{\omega}^{ai}(\Sigma)\equiv\frac{\delta \Sigma}{\delta \omega_i^a} -\partial_{\mu}\frac{%
\delta\Sigma}{\delta\rho^{a}_{\mu i}} -igf^{abc}\bar{\omega}_i^b%
\frac{\delta \Sigma}{\delta b^c} -gf^{abc}A_{\mu}^b\frac{\delta
\Sigma}{\delta {\mathcal{C}}_{\mu i}^c} = \Delta^{ai}_{\omega}\;,
\label{eqmotion-omega}
\end{align}
and
\begin{align}
\Delta^{ai}_{\omega}=-\partial^{2}\bar\omega^{a}_{i}+\partial_{\mu}\bar%
\rho^{a}_{\mu i} + \partial_{\mu} \bar{\mathcal{C}}^a_{\mu i} \;. \label{delta-omega}
\end{align}}
Notice that all breaking terms in eqs.\eqref{delta-baromega},  \eqref{delta-varphi}, \eqref{delta-omega} are linear.

\item{The integrated Ward identities involving the auxiliary and the Faddeev-Popov ghost fields:
\begin{equation}
\mathcal{I}^{\,i}(\Sigma)\equiv\int d^{4}x\;\left( c^{a}\frac{\delta \Sigma}{\delta \omega
_{i}^{a}}-\bar{\omega}_{i}^{a}\frac{\delta \Sigma}{\delta \bar{c}%
^{a}}
-(\partial_{\mu}c^{a})\frac{\delta\Sigma}{\delta\mathcal{C}^{a}_{\mu i}}
+\frac{\delta \Sigma}{\delta \mathcal{C}_{\mu i}^{c}}\frac{%
\delta \Sigma}{\delta \Omega _{\mu }^{c}}\right) =0\;,\label{int1}
\end{equation}
\begin{equation}
\mathcal{J}^{\,i}(\Sigma)\equiv\int d^{4}x\left( c^{a}\frac{\delta \Sigma }{\delta \varphi _{i}^{a}}+\bar{%
\varphi}_{i}^{a}\frac{\delta \Sigma }{\delta \bar{c}^{a}}-\frac{\delta
\Sigma }{\delta {L}^{a}}\frac{\delta \Sigma }{\delta \omega _{i}^{a}}-\frac{%
\delta \Sigma }{\delta \Omega _{\mu }^{a}}\frac{\delta \Sigma }{\delta \eta
_{\mu i}^{a}}+\gamma ^{2}\delta _{\mu a}^{i}\frac{\delta \Sigma }{\delta
\Omega _{\mu }^{a}}\right) =0\;.\label{int2}
\end{equation}}

\item{The linearly broken ghost Ward identity  \cite{Blasi:1990xz,Piguet:1995er}:
\begin{equation}
\mathcal{G}^{a}(\Sigma)=\Delta _{c}^{a}\;, \label{ghward}
\end{equation}
where
\begin{eqnarray}
\mathcal{G}^{a} &=&\int d^{4}x\biggl(\frac{\delta }{\delta {c}^{a}}
-igf^{abc}\bar{c}^{b}\frac{\delta }{\delta b^{c}}+gf^{abc}\bar{\omega}_{i}^{b}\frac{%
\delta }{\delta \bar{\varphi}_{i}^{c}}+gf^{abc}\varphi _{i}^{b}\frac{\delta }{\delta
\omega _{i}^{c}}+gf^{abc}\bar{\eta}_{\mu i}^{b}\frac{\delta }{\delta \rho _{i}^{c}}\nonumber
\\
&&{\ }{\ }{\ }{\ }+gf^{abc}\bar{\mathcal{C}}_{\mu i}^{b}\frac{\delta }{\delta \lambda _{\mu i}^{c}}
+gf^{abc}\bar{\rho}_{\mu i}^{b}\frac{\delta }{\delta \bar{\lambda}_{\mu i}^{c}}
+gf^{abc}\eta_{\mu i}^{b}\frac{\delta }{\delta \mathcal{C}_{\mu i}^{c}}\biggr)\;,
\end{eqnarray}
and the linear breaking $\Delta _{c}^{a}$ given by
\begin{equation}
\Delta _{c}^{a}=\int d^{4}x\,gf^{abc}(\Omega _{\mu }^{b}A_{\mu
}^{c}-L^{b}c^{c}-\gamma ^{2}\delta _{c\mu }^{i}\bar{\rho}_{\mu i}^{b})\;.
\end{equation}}

\item{The linearly broken identity involving only the auxiliary fields:
\begin{equation}
\mathcal{N}_{ij}\left( \Sigma\right) =-\gamma ^{2}\int
d^{4}x\;\delta ^{ab}\delta _{\mu \nu }\delta _{\nu b}^{j}\left( {\bar{\rho}}%
_{\mu i}^{a}+\bar{\mathcal{C}}_{\mu i}^{a}\right) \;,  \label{nij}
\end{equation}%
where
\begin{equation}
\mathcal{N}_{ij}=\int d^{4}x\left( -{\bar{\omega}}_{i}^{a}\frac{\delta }{%
\delta {\bar{\varphi}}_{j}^{a}}+{\varphi }_{j}^{a}\frac{\delta }{\delta {%
\omega }_{i}^{a}}+{\bar{\eta}}_{\mu j}^{a}\frac{\delta }{\delta {\rho }_{\mu
i}^{a}}-{\bar{\mathcal{C}}}_{\mu i}^{a}\frac{\delta }{\delta {\lambda }_{\mu
j}^{a}}-({\bar{\rho}_{\mu i}^{a}}+{\bar{\mathcal{C}}}_{\mu i}^{a})\frac{%
\delta }{\delta {\bar{\lambda}}_{\mu j}^{a}}+({\eta }_{\mu j}^{a}+{\bar{\eta}%
}_{\mu j}^{a})\frac{\delta }{\delta \mathcal{C}_{\mu i}^{a}}\right)\;.
\label{nijop}
\end{equation}}

\item The linearly broken global symmetry $U(f)$, $f=4(N^2-1)$:
\begin{equation}
\mathcal{Q}_{ij}(\Sigma)=\gamma ^{2}\delta ^{ab}\delta _{\mu \nu
}\int d^{4}x(\delta _{b\nu }^{j}\eta _{\mu i}^{a}-\delta _{b\nu }^{i}\bar{%
\lambda}_{\mu j}^{a})\;, \label{Qij}
\end{equation}%
where
\begin{align}
\mathcal{Q}_{ij}& =\int d^{4}x \left( \varphi _{i}^{a}\frac{\delta }{\delta
\varphi _{j}^{a}}-\bar{\varphi}_{j}^{a}\frac{\delta }{\delta \bar{\varphi}%
_{i}^{a}}+\omega _{i}^{a}\frac{\delta }{\delta \omega _{j}^{a}}-\bar{\omega}%
_{i}^{a}\frac{\delta }{\delta \bar{\omega}_{j}^{a}}+\mathcal{C}_{\mu i}^{a}%
\frac{\delta }{\delta \mathcal{C}_{\mu j}^{a}}-\bar{\mathcal{C}}_{\mu j}^{a}%
\frac{\delta }{\delta \bar{\mathcal{C}}_{\mu i}^{a}}\right.  \notag \\
& \left. +\rho _{\mu i}^{a}\frac{\delta }{\delta \rho _{\mu j}^{a}}-\bar{\rho%
}_{\mu j}^{a}\frac{\delta }{\delta \bar{\rho}_{\mu i}^{a}}+\eta _{\mu i}^{a}%
\frac{\delta }{\delta \eta _{\mu j}^{a}}-\lambda _{\mu j}^{a}\frac{\delta }{%
\delta \lambda _{\mu i}^{a}}+\bar{\eta}_{\mu i}^{a}\frac{\delta }{\delta
\bar{\eta}_{\mu j}^{a}}-\bar{\lambda}_{\mu j}^{a}\frac{\delta }{\delta \bar{%
\lambda}_{\mu i}^{a}}\right) \;.  \label{uf}
\end{align}

\item The linearly broken rigid identity:
\begin{equation}
\mathcal{W}^{a}(\Sigma)=\gamma^{2}f^{abc}\int d^{4}x\left(
{\bar{\lambda}}_{\mu \mu }^{bc}+{\eta }_{\mu \mu }^{bc}\right) \;,
\label{color}
\end{equation}
with
\begin{eqnarray}
\mathcal{W}^{a} &=&\int d^{4}x\;f^{abc}\Bigl(A_{\mu }^{b}\frac{\delta }{%
\delta A_{\mu }^{c}}+\Omega _{\mu }^{b}\frac{\delta }{\delta \Omega _{\mu
}^{c}}+c^{b}\frac{\delta }{\delta c^{c}}+L^{b}\frac{\delta }{\delta L^{c}}+{%
\bar{c}}^{b}\frac{\delta }{\delta {\bar{c}}^{c}}+b^{b}\frac{\delta }{\delta
b^{c}}  \notag \\
&&\left. +{\bar{\omega}}_{i}^{b}\frac{\delta }{\delta {\bar{\omega}}_{i}^{c}}%
+{\ \omega }_{i}^{b}\frac{\delta }{\delta {\omega }_{i}^{c}}+{\bar{\varphi}}%
_{i}^{b}\frac{\delta }{\delta {\bar{\varphi}}_{i}^{c}}+{\varphi }_{i}^{b}%
\frac{\delta }{\delta {\varphi }_{i}^{c}}+{\bar{\eta}}_{\mu i}^{b}\frac{%
\delta }{\delta {\bar{\eta}}_{\mu i}^{c}}+{\ \eta }_{\mu i}^{b}\frac{\delta
}{\delta {\ \eta }_{\mu i}^{c}}+{\bar{\mathcal{C}}}_{\mu i}^{b}\frac{\delta
}{\delta {\bar{\mathcal{C}}}_{\mu i}^{c}}\right. \nonumber \\
&&+{{\mathcal{C}}}_{\mu i}^{b}\frac{\delta }{\delta {{\mathcal{C}}}_{\mu
i}^{c}}+{\bar{\rho}}_{\mu i}^{b}\frac{\delta }{\delta {\bar{\rho}}_{\mu
i}^{c}}+{\rho }_{\mu i}^{b}\frac{\delta }{\delta {\rho }_{\mu i}^{c}}+{\bar{%
\lambda}}_{\mu i}^{b}\frac{\delta }{\delta {\bar{\lambda}}_{\mu i}^{c}}+{%
\lambda }_{\mu i}^{b}\frac{\delta }{\delta {\lambda }_{\mu i}^{c}}\Bigl)\;. \label{Rigid}
\end{eqnarray}
\end{itemize}

%%%%%%%%%%%%%%%%%%%
\section{Stability and invariant counterterm}
%%%%%%%%%%%%%%%%%%%
%%%%%%%%%%%%%%%%%%%%%%%%%%%%%%%%%%
\subsection{Algebraic characterization of the invariant counterterm}
%%%%%%%%%%%%%%%%%%%%%%%%%%%%%%%%%%
In order to characterize the most general invariant counterterm which can be freely added to all orders
in perturbation theory we follow the set up of the Algebraic Renormalization  \cite{Piguet:1995er} and perturb the classical action $\Sigma$ by adding an integrated local polynomial in the fields and sources, $\Sigma^{ct}$, with dimension bounded by four, and with vanishing ghost number. We demand thus that the perturbed action, $(\Sigma +\epsilon\Sigma^{ct})$, where $\epsilon$ is an expansion parameter, fulfills, to the first order in $\epsilon$, the same Ward identities obeyed by the classical action $\Sigma$, {\it i.e.} equations \eqref{st1} -- \eqref{color}. This requirement gives rise to the following constraints for the counterterm $\Sigma^{ct}$:
\begin{eqnarray}
&\mathcal{B}_{\Sigma}\Sigma^{ct}=0\,,&\label{linST}\\\nonumber\\
&\displaystyle\frac{\delta\Sigma^{ct}}{\delta b^{a}}=0\,,\qquad
\frac{\delta\Sigma^{ct}}{\delta\eta^{a}_{\mu i}}=0\,,\qquad
\frac{\delta\Sigma^{ct}}{\delta\bar\lambda^{a}_{\mu i}}=0\,,\qquad
\frac{\delta\Sigma^{ct}}{\delta \mathcal{C}^{a}_{\mu i}}=0\,,\qquad
\frac{\delta\Sigma^{ct}}{\delta\bar\rho^{a}_{\mu i}}=0\,,\qquad
\frac{\partial\Sigma^{ct}}{\partial\gamma^{2}}=0\,,&\label{CTindep}\\\nonumber\\
&\displaystyle\frac{\delta\Sigma^{ct}}{\delta\bar{c}^{a}}+\partial_{\mu}\frac{\delta\Sigma^{ct}}{\delta \Omega^{a}_{\mu}}=0\,,&\label{antighost}\\\nonumber\\
&\Xi^{ai}_{\bar\varphi}(\Sigma^{ct})=0\,,\qquad
\Xi^{ai}_{\omega}(\Sigma^{ct})=0\,,&\\\nonumber\\
&{\Xi^{ai}_{\varphi}}_{\Sigma}(\Sigma^{ct})=0\,,\qquad
{\Xi^{ai}_{\bar\omega}}_{\Sigma}(\Sigma^{ct})=0\,,\qquad
\mathcal{I}^{i}_{\Sigma}(\Sigma^{ct})=0\,,\qquad
\mathcal{J}^{i}_{\Sigma}(\Sigma^{ct})=0\,,&\label{linops}\\\nonumber\\
&\mathcal{G}^{a}(\Sigma^{ct})=0\,,\qquad
\mathcal{N}_{ij}(\Sigma^{ct})=0\,,\qquad
\mathcal{Q}_{ij}(\Sigma^{ct})=0\,,\qquad
\mathcal{W}^{a}(\Sigma^{ct})=0\,.&\label{Qcharge}
\end{eqnarray}
Equations \eqref{CTindep} imply that the counterterm is independent from the fields $(b^{a},\eta^{a}_{\mu i},\bar\lambda^{a}_{\mu i},\mathcal{C}^{a}_{\mu i},\bar\rho^{a}_{\mu i})$ and from the parameter $\gamma^{2}$. From equation ~\eqref{antighost} it follows that  $\bar{c}^{a}$ and $\Omega^{a}_{\mu}$ enter  only through the combination $(\partial_{\mu}\bar{c}^{a}+\Omega^{a}_{\mu})$. We also notice that the functional operators in equations~\eqref{linops} are those associated with the Ward identities \eqref{eqmotion-baromega}, \eqref{eqmotion-phi}, \eqref{int1} and \eqref{int2},  namely
\begin{eqnarray}
{\Xi^{ai}_{\bar\omega}}_{\Sigma}&=&\frac{\delta}{\delta \bar{\omega}_i^a} +\partial_{\mu}
\frac{\delta}{\delta \bar{\mathcal{C}}_{\mu i}^a}
+gf^{abc}A^{b}_{\mu}\frac{\delta}{\delta\bar\rho^{c}_{\mu i}
} + gf^{abc}\left(\frac{\delta\Sigma}{\delta\bar\lambda^{b}_{\mu
i}}-\gamma^{2}\delta^{i}_{\mu b}\right) \frac{\delta}{
\delta\Omega^{c}_{\mu}}-gf^{abc}\frac{\delta\Sigma}{
\delta\Omega^{b}_{\mu}}\frac{\delta}{
\delta\bar\lambda^{c}_{\mu i}}\,,\\\cr
{\Xi^{ai}_{\varphi}}_{\Sigma}&=&\frac{\delta}{\delta \varphi _{i}^{a}}-\partial _{\mu }
\frac{\delta }{\delta \bar{\eta}_{\mu i}^{a}}-igf^{abc}
\bar{\varphi}_{i}^{b}\frac{\delta }{\delta b^{c}}+gf^{abc}
\bar{\omega}_{i}^{b}\frac{\delta }{\delta \bar{c}^{c}}
-gf^{abc}A_{\mu }^{b}\frac{\delta }{\delta \eta _{\mu
i}^{c}}\nonumber\\
&&-gf^{abc}\frac{\delta \Sigma}{\delta \mathcal{C}_{\mu
i}^{b}}\frac{\delta}{\delta \Omega _{\mu }^{c}}
-gf^{abc}\frac{\delta \Sigma}{\delta \Omega _{\mu }^{b}}
\frac{\delta}{\delta \mathcal{C}_{\mu
i}^{c}}\,,\\\cr
\mathcal{I}^{\,i}_\Sigma&=&\int d^{4}x\;\left( c^{a}\frac{\delta}{\delta \omega
_{i}^{a}}-\bar{\omega}_{i}^{a}\frac{\delta }{\delta \bar{c}^{a}}
-(\partial_{\mu}c^{a})\frac{\delta}{\delta\mathcal{C}^{a}_{\mu i}}
+\frac{\delta \Sigma}{\delta \mathcal{C}_{\mu i}^{a}}
\frac{\delta }{\delta \Omega _{\mu }^{a}}
-\frac{\delta \Sigma}{\delta \Omega _{\mu }^{a}}
\frac{\delta }{\delta \mathcal{C}_{\mu i}^{a}}\right)\,,\\\cr
\mathcal{J}^{\,i}_\Sigma&=&\int d^{4}x\left( c^{a}\frac{\delta}{\delta \varphi _{i}^{a}}+\bar{
\varphi}_{i}^{a}\frac{\delta }{\delta \bar{c}^{a}}
+\gamma ^{2}\delta _{\mu a}^{i}\frac{\delta }{\delta
\Omega _{\mu }^{a}}
-\frac{\delta\Sigma }{\delta {L}^{a}}\frac{\delta  }{\delta \omega _{i}^{a}}
-\frac{\delta \Sigma }{\delta \omega _{i}^{a}}\frac{\delta }{\delta {L}^{a}}
-\frac{\delta \Sigma }{\delta \Omega _{\mu }^{a}}\frac{\delta  }{\delta\eta_{\mu i}^{a}}
-\frac{\delta \Sigma }{\delta\eta_{\mu i}^{a}}\frac{\delta }{\delta \Omega _{\mu }^{a}}\right)\,.
\end{eqnarray}
Further, the trace of the operator $\mathcal{Q}_{ij}$ in eq.\eqref{Qcharge}, defines a new quantum number, the $Q$-charge, displayed in  table \ref{table}. Finally, reminding that the operator $\mathcal{B}_{\Sigma}$ is nilpotent, eq.\eqref{nilp1},  from eq.~\eqref{linST} it follows that  $\Sigma^{ct}$ can be characterized by looking at the cohomology of the operator $\mathcal{B}_{\Sigma}$ in the space of the integrated local polynomials with dimension four and zero ghost number and $Q$-charge. Thus, according to the general results on the cohomology of Yang-Mills theories \cite{Piguet:1995er},  we start by writing the most general invariant counterterm as
\begin{equation}
\Sigma ^{ct}=\frac{a_{0}}{4}\int d^{4}x\, F_{\mu \nu }^{a}F_{\mu \nu }^{a}+%
\mathcal{B}_{\Sigma }\Delta ^{(-1)}\;.  \label{Contra}
\end{equation}
The first term in the r.h.s of  equation \eqref{Contra}  represents the nontrivial part of the cohomology of $\mathcal{B}_{\Sigma}$, with $a_0$ being an arbitrary coefficient, while $\Delta ^{(-1)}$ is an integrated polynomial in the fields and sources with dimension $4$, ghost number $-1$ and vanishing $Q$-charge.

\begin{table}
\begin{tabular}{|l|c|c|c|c|c|c|c|c|c|c|c|c|c|c|c|c|c|c|}
\hline
& $A$ & $b$ & $c$ & $\bar{c}$ & $\Omega $ & $L$ & $\varphi $ & $\bar{\varphi}
$ & $\omega $ & $\bar{\omega}$ & $\lambda $ & $\bar{\lambda}$ & $\eta $ & $%
\bar{\eta}$ & $\mathcal{C}$ & $\mathcal{\bar{C}}$ & $\rho $ & $\,\bar{\rho}\phantom{\Bigl|}$
\\ \hline
dimension & 1 & 2 & 0 & 2 & 3 & 4 & 1 & 1 & 1 & 1 & 2 & 2 & 2 & 2 & 2 & 2 & 2 & 2
\\ \hline
ghost number
& 0 & 0 & 1 & $-1$ & $-1$ & $-2$ & 0 & 0 & 1 & $-1$ & 0 & 0 & 0 & 0 & 1 & $-1$ & 1 &
$-1$ \\ \hline
$Q$-charge & 0 & 0 & 0 & 0 & 0 & 0 & 1 & $-1$ & 1 & $-1$ & $-1$ & $-1$ & 1 & 1 & 1 &
$-1$ & 1 & $-1$ \\ \hline
\end{tabular}
\caption{The quantum numbers of fields and sources of the theory}
\label{table}
\end{table}
\noindent By using the equations~\eqref{CTindep} and \eqref{antighost}, we find
\begin{eqnarray}
\Delta ^{(-1)} &=&\int d^{4}x\left( a_{1}\,A_{\mu }^{a}(\Omega _{\mu
}^{a}+\partial _{\mu }\bar{c}^{a})+a_{2}\,L^{a}c^{a}+a_{3}\,gf^{abc}A_{\mu
}^{a}\varphi _{i}^{b}\mathcal{\bar{C}}_{\mu i}^{c}+a_{4}\,\varphi
_{i}^{a}\partial _{\mu }\mathcal{\bar{C}}_{\mu i}^{a}+a_{5}\,gf^{abc}A_{\mu
}^{a}\bar{\omega}_{i}^{b}\bar{\eta}_{\mu i}^{c}\right. \nonumber \\
&&+a_{6}\,\bar{\omega}_{i}^{a}\partial _{\mu }\bar{\eta}_{\mu i}^{a}+a_{7}\,
\mathcal{\bar{C}}_{\mu i}^{a}\bar{\eta}_{\mu i}^{a}
+a_{8}\,\bar{\omega}_{i}^{a}\partial ^{2}\varphi
_{i}^{a}
+a_{9}\,gf^{abc}\bar{\omega}_{i}^{a}A_{\mu }^{c}\partial _{\mu
}\varphi _{i}^{b}
+a_{10}\,gf^{abc}A_{\mu }^{c}\partial _{\mu }\bar{\omega}%
_{i}^{a}\varphi _{i}^{b}\nonumber\\
&&\left.
+\beta_{1}^{abcd}\,\varphi
_{i}^{a}\bar\varphi _{i}^{b}\varphi _{j}^{c}\bar{\omega}_{j}^{d}
+\beta_{2}^{abcd}\,\varphi
_{i}^{a}\bar\varphi _{j}^{b}\varphi _{i}^{c}\bar{\omega}_{j}^{d}
+\beta_{3}^{abcd}\,\omega_{i}^{a}\bar{\omega}_{i}^{b}{\varphi}_{j}^{c}\bar{\omega}%
_{j}^{d}
+\beta_{4}^{abcd}\,\omega_{i}^{a}\bar{\omega}_{j}^{b}{\varphi}_{i}^{c}\bar{\omega}%
_{j}^{d}\right)\,, \label{Delta}
\end{eqnarray}
where $a_k$ $(k=1,\dots,10)$ are constant arbitrary coefficients, while $\beta^{abcd}_{l}$ $(l=1,2,3,4)$ stands for an  invariant tensor of rank 4. Using the remaining constrains on $\Sigma^{ct}$, and taking into account the following useful relations:
\begin{equation}
\left[\mathcal{B}_{\Sigma},\Xi^{ai}_{\bar\varphi}\right]=-{\Xi^{ai}_{\bar\omega}}_{\Sigma}\,,\qquad
\left\{\mathcal{B}_{\Sigma},\Xi^{ai}_{\omega}\right\}={\Xi^{ai}_{\varphi}}_{\Sigma}\,,\qquad
\left[\mathcal{B}_{\Sigma},\mathcal{I}^{i}_{\Sigma}\right]=-\mathcal{J}^{i}_{\Sigma}\,,\qquad
\left\{\mathcal{B}_{\Sigma},\mathcal{G}^{a}\right\}=\mathcal{W}^{a}\,,
\end{equation}
it turns out that $\Delta ^{(-1)}$ depends only from one free parameter $a_1$, namely
\begin{equation}
\Delta ^{(-1)}=a_{1}\int d^{4}x\left( \left( \partial _{\mu }\bar{c}%
^{a}+\Omega _{\mu }^{a}\right) A_{\mu }^{a}+\bar{\omega}_{i}^{a}\partial
_{\mu }D_{\mu }^{ab}\varphi _{i}^{b}+\bar{\eta}_{\mu i}^{a}D_{\mu }^{ab}\bar{%
\omega}_{i}^{b}-\mathcal{\bar{C}}_{\mu i}^{a}D_{\mu }^{ab}\varphi _{i}^{b}+%
\mathcal{\bar{C}}_{\mu i}^{a}\bar{\eta}_{\mu i}^{a}\,\right)\,.
\label{Deltafinal}
\end{equation}
Therefore, for the final form of the most general allowed invariant counterterm one finds
\begin{eqnarray}
\Sigma^{ct} &=&\frac{a_{0}}{4}\int d^{4}x\,F_{\mu \nu }^{a}F_{\mu \nu
}^{a}+a_{1}\int d^{4}x\left( A_{\mu }^{a}\,D^{ab}_{\nu}F^{b}_{\mu\nu}
+\partial _{\mu }c^{a}\partial _{\mu }\bar{c}^{a}+\partial _{\mu
}c^{a}\Omega _{\mu }^{a}+gf^{bac}\partial _{\mu }c^{a}\partial _{\mu }\bar{%
\omega}_{i}^{c}\varphi _{i}^{b}\right.  \nonumber\\
&&\left. -gf^{mab}\partial _{\mu }c^{a}\bar{\eta}_{\mu i}^{m}\bar{\omega}%
_{i}^{b}+gf^{mab}\partial _{\mu }c^{a}\mathcal{\bar{C}}_{\mu i}^{m}\varphi
_{i}^{b}-\lambda _{\mu i}^{a}\partial _{\mu }\varphi _{i}^{a}+\lambda _{\mu
i}^{a}\bar{\eta}_{\mu i}^{a}+\rho _{\mu i}^{a}\partial _{\mu }\bar{\omega}%
_{i}^{a}-\mathcal{\bar{C}}_{\mu i}^{a}\rho _{\mu i}^{a}\right.  \notag \\
&&\left. -\bar{\varphi}_{i}^{a}\partial _{\mu }\bar{\eta}_{\mu i}^{a}+\omega
_{i}^{a}\partial _{\mu }\mathcal{\bar{C}}_{\mu i}^{b}+\bar{\varphi}%
_{i}^{a}\partial ^{2}\varphi _{i}^{a}-\bar{\omega}_{i}^{a}\partial
^{2}\omega _{i}^{a}\right)\,.  \label{contrafinal}
\end{eqnarray}

%%%%%%%%%%%%%%%%%%%%%%%%%%%%%%%%%%%%%%%%
\subsection{Renormalization and $Z$-factors}
%%%%%%%%%%%%%%%%%%%%%%%%%%%%%%%%%%%%%%%%

It remains to check that the counterterm \eqref{contrafinal}
can be reabsorbed in the starting action $\Sigma$ by means of a multiplicative redefinition of the fields, sources and parameters of the theory, {\it i.e.}
\begin{equation}
\Sigma(\phi,\Phi )+\epsilon\,\Sigma ^{ct}(\phi,\Phi)=\Sigma(\phi_0,\Phi _{0})+O(\epsilon ^{2})
\label{renorm}
\end{equation}
where $(\phi,\phi_0)$ denote  the renormalized and bare fields, respectively, while $(\Phi,\Phi_0)$ stand for the renormalized and bare sources and parameters, {\it i.e.} $\Phi= (L^{a},\Omega^{a}_{\mu},g,\gamma^{2})$. The renormalized and bare quantities are related each other as
\begin{equation}
\phi_{0}=Z_{\phi}^{1/2}\,\phi\,,\qquad
\Phi _{0}=Z_{\Phi }\,\Phi\,,
\end{equation}
where the $Z$'s are renormalization factors. One can easily proove that eq.\eqref{renorm} is in fact fulfilled by writing the bare action in terms of two independent renormalization factors $Z_g$ and $Z_A^{1/2}$, {\it i.e.}:
\begin{eqnarray}
\Sigma(\phi_0,\Phi_0)&=&\int d^{4}x\left( \frac{Z_{A}}{2}(\partial _{\mu }A_{\nu
}^{a})\partial _{\mu }A_{\nu }^{a}
+Z_{g}Z_{A}^{3/2}gf^{abc}(\partial _{\mu
}A_{\nu }^{a})A_{\mu }^{b}A_{\nu }^{c}
+\frac{1}{4}Z_{g}^{2}Z_{A}^{2}g^{2}f^{abc}f^{ade}A_{\mu }^{b}A_{\nu }^{c}A_{\mu
}^{d}A_{\nu }^{e}\right.\nonumber\\
&&+iZ_{A}^{1/4}b^{a}\partial _{\mu }A_{\mu }^{a}
+Z_{g}^{-1}Z_{A}^{-1/2}\bar{c}^{a}\partial^{2}c^{a}
+gf^{abc}(\partial _{\mu }\bar{c}^{a}) A_{\mu }^{c}c^{b}\nonumber \\
&&-Z_{g}^{-1}Z_{A}^{-1/2}\bar{\varphi}_{i}^{a}\partial ^{2}\varphi_{i}^{a}
-gf^{abc}(\partial _{\mu}\bar{\varphi}_{i}^{a})A_{\mu }^{c}\varphi _{i}^{b}
+Z_{g}^{-1}Z_{A}^{-1/2}\bar{\omega}_{i}^{a}\partial ^{2}\omega _{i}^{a} \nonumber\\
&&+gf^{abc}(\partial_{\mu}\bar{\omega}_{i}^{a})A_{\mu }^{c}\omega_{i}^{b}
+Z_{g}^{-1}Z_{A}^{-1/2}gf^{amb}(\partial _{\nu }\bar{\omega
}_{i}^{a})\partial _{\nu }c^{m}\varphi_{i}^{b}
+g^{2}f^{amb}f^{mnp}(\partial _{\nu }\bar{\omega}_{i}^{a})A_{\mu }^{n}c^{p}\varphi _{i}^{b} \nonumber\\
&&+Z_{g}^{-1}Z_{A}^{-1/2}\lambda _{\mu i}^{a}\partial _{\mu }\varphi_{i}^{a}
+gf^{acb}\lambda _{\mu i}^{a}A_{\mu}^{c}\varphi _{i}^{b}
-Z_{g}^{-1}Z_{A}^{-1/2}\bar{\eta}_{\mu i}^{a}\partial_{\mu }\bar{\varphi}_{i}^{a} \nonumber\\
&&-gf^{acb}\bar{\eta}_{\mu i}^{a}A_{\mu }^{c}\bar{\varphi}_{i}^{b}
+Z_{g}^{-1}Z_{A}^{-1/2}gf^{abc}\bar{\eta}_{\mu i}^{c}(\partial _{\mu }c^{a})\bar{\omega}_{i}^{b}
+g^{2}f^{abc}f^{amp}\bar{\eta}_{\mu
i}^{c}A_{\mu }^{m}c^{p}\bar{\omega}_{i}^{b} \nonumber\\
&&-Z_{g}^{-1}Z_{A}^{-1/2}\lambda _{\mu i}^{a}\bar{\eta}_{\mu
i}^{c}
-Z_{g}^{-1}Z_{A}^{-1/2}\mathcal{\bar{C}}_{\mu i}^{a}\partial_{\mu}\omega_{i}^{a}
-gf^{acb}\mathcal{\bar{C}}_{\mu i}^{a}A_{\mu}^{c}\omega _{i}^{b}
-Z_{A}^{-1/2}Z_{g}^{-1}\rho _{\mu i}^{a}\partial _{\mu }\bar{\omega}_{i}^{a} \nonumber\\
&&-gf^{acb}\rho _{\mu i}^{a}A_{\mu }^{c}\bar{\omega}_{i}^{b}
+\mathcal{\bar{C}}_{\mu i}^{c}\mathcal{C}_{\mu i}^{c}
+Z_{g}^{-1}Z_{A}^{-1/2}\mathcal{\bar{C}}_{\mu i}^{c}\rho _{\mu i}^{c}
+Z_{g}^{-1}Z_{A}^{-1/2}gf^{abc}\mathcal{\bar{C}}_{\mu i}^{c}(\partial_{\mu}c^{a})\varphi _{i}^{b} \nonumber\\
&&+g^{2}f^{amp}f^{abc}\mathcal{\bar{C}}_{\mu
i}^{c}A_{\mu }^{m}c^{p}\varphi _{i}^{b}
-\rho _{\mu i}^{a}\bar{\rho}_{\mu
i}^{a}
+\gamma ^{2}\bar{\lambda}_{\mu i}^{a}\delta
^{ab}\delta _{\mu \nu }\delta _{\nu b}^{i}
-\bar{\lambda}_{\mu i}^{a}\bar{\eta}_{\mu
i}^{a}+\gamma ^{2}\eta _{\mu i}^{a}\delta ^{ab}\delta
_{\mu \nu }\delta _{\nu b}^{i}\nonumber\\
&&\left.
-\eta _{\mu i}^{a}\lambda _{\mu i}^{a}
-Z_{g}^{-1}Z_{A}^{-1/2}\Omega_{\mu }^{a}\partial_{\mu }c^{a}
-g\Omega _{\mu}^{a}f^{acb}A_{\mu }^{c}c^{c}
+\frac{g}{2}f^{abc}L^{a}c^{b}c^{c}\right)\,.
\end{eqnarray}
The two independent renormalization factors $Z_g$ and $Z^{1/2}_A$  are related to the coefficients $a_0$ and $a_1$ in the following way:
\begin{eqnarray}
Z_{g} &=&1-\frac{\epsilon }{2}a_{0}\,,\nonumber \\
Z_{A}^{1/2} &=&1+\epsilon \left( \frac{a_{0}}{2}+a_{1}\right)\,. \\
\end{eqnarray}
All remaining renormalization factors can be expressed in terms of $Z_A^{1/2}$ and $Z_g$ as
\begin{eqnarray}
&Z_{b} =Z_{A}^{-1/2}\,,\qquad Z_{c}=Z_{\bar{c}}=Z_{\varphi }=Z_{\bar{%
\varphi}}=Z_{\lambda }=Z_{\bar{\eta}}=Z_{g}^{-1}Z_{A}^{-1/2}\,,& \nonumber\\
&Z_{\bar{\omega}}^{1/2} =Z_{\mathcal{\bar{C}}}^{1/2}=Z_{g}^{-1}\,,\qquad
Z_{\rho }^{1/2}=Z_{\omega }^{1/2}=Z_{A}^{-1/2}\,,\qquad
Z_{\Omega }=Z_{g}^{-1/2}Z_{A}^{-1/4}\,,\qquad Z_{\mathcal{C}}^{1/2}=Z_{g}\,,&
\nonumber\\
&Z_{L} =Z_{\bar{\rho}}^{1/2}=Z_{A}^{1/2}\,,\qquad
Z_{\eta }=Z_{\bar{\lambda}}=Z_{g}Z_{A}^{1/2}\,,\qquad
Z_{\gamma ^{2}}=Z_{g}^{-1/2}Z_{A}^{-1/4}\,.&
\end{eqnarray}
In particular, we point out the two relations: $Z_c Z_A^{1/2} Z_g=1$ and $Z_{\gamma ^{2}}=Z_{g}^{-1/2}Z_{A}^{-1/4}$. The first one expresses the nonrenormalization theorem of the ghost-antighost-gluon vertex. The second relation tells us that the renormalization factor $Z_{\gamma^2}$ is not an independent parameter of the theory. Within the linear breaking formulation, this result is a consequence of the Ward identity \eqref{nonrenorm}. This ends the proof of the multiplicative renormalizability of the Gribov-Zwanziger action in the linear breaking formulation

%%%%%%%%%%%%%%%%%%%%%%%%%%%%%%%%%%%%%%%%%%%%%%%%%%
\section{Generalization to the Refined Gribov-Zwanziger}
%%%%%%%%%%%%%%%%%%%%%%%%%%%%%%%%%%%%%%%%%%%%%%%%%%

The results established in the previous section can be generalized to the so-called Refined Gribov-Zwanziger action  \cite{Dudal:2007cw,Dudal:2008sp}, which takes into account the effects of the dimension two condensates $\langle A^a_\mu A^a_\mu \rangle$ and $\langle {\bar \varphi}^{a}_i \varphi^{a}_i -   {\bar \omega}^{a}_i \omega^{a}_i \rangle$.  The presence of these condensates can  be encoded in the starting action by modifying the Gribov-Zwanziger action as
\begin{equation}
S_{RGZ}=S_{GZ}+\mu ^{2}\int d^{4}x\left( \bar{\varphi}_{i}^{a}\varphi
_{i}^{a}-\bar{\omega}_{i}^{a}\omega _{i}^{a}\right) +\frac{m^{2}}{2}\int d^{4}x\,
A^{a}_{\mu}A^{a}_{\mu}\,, \label{RGZ}
\end{equation}
where the mass parameters $\mu,m$ are dynamical parameters related to the aforementioned dimension two condensates, see  \cite{Vandersickel:2011ye} for an updated discussion.  The Refined Gribov-Zwanziger action gives rise to a gluon propagator which remains suppressed in IR region. However, unlike the propagator of the Gribov-Zwanziger theory, eq.\eqref{gribprop}, it  attains a non-vanishing value at the origin in momentum space
\begin{equation}
\left\langle A_{\mu }^{a}(k)A_{\nu }^{b}(-k)\right\rangle _{RGZ}=\delta ^{ab}%
\frac{k^{2}+\mu ^{2}}{k^{4}+(m^{2}+\mu ^{2})k^{2}+m^{2}\mu ^{2}+\gamma ^{4}}%
\left( \delta _{\mu \nu }-\frac{k_{\mu }k_{\nu }}{k^{2}}\right)\,, \label{rgzgluon}
\end{equation}
Expression \eqref{rgzgluon} in in agreement with the recent lattice simulations  \cite{Cucchieri:2007rg,Cucchieri:2008fc,Bornyakov:2009ug,Dudal:2010tf}. A similar behavior for the gluon propagator has emerged from the studies of the Schwinger-Dyson equations  \cite{Aguilar:2008xm,Fischer:2008uz}, see also \cite{Cornwall:1981zr} for the introduction of the concept of the dynamical gluon mass. \\\\In order to discuss the Refined Gribov-Zwanziger action, eq.\eqref{RGZ}, within the linearly broken BRST formulation, we analyze the two terms,  $\left( \bar{\varphi}_{i}^{a}\varphi_{i}^{a}-\bar{\omega}_{i}^{a}\omega _{i}^{a}\right)$ and $A^{a}_{\mu}A^{a}_{\mu}$, separately. Let us start with $\left( \bar{\varphi}_{i}^{a}\varphi_{i}^{a}-\bar{\omega}_{i}^{a}\omega _{i}^{a}\right)$, which is BRST invariant. In fact
\begin{equation}
\int d^{4}x\left( \bar{\varphi}_{i}^{a}\varphi _{i}^{a}-\bar{\omega}%
_{i}^{a}\omega _{i}^{a}\right) =s\int d^{4}x\left( \bar{\varphi}%
_{i}^{a}\omega _{i}^{a}\right)\,.
\end{equation}
This term will thus not affect the BRST breaking. Moreover, it is easy to check that all Ward identities in eqs.\eqref{st1}, \eqref{eqmotion1}, \eqref{eqmotion2}, \eqref{eqmotion3}, \eqref{nonrenorm},  \eqref{eqmotion-barphi}, \eqref{eqmotion-baromega}, \eqref{eqmotion-phi}, \eqref{eqmotion-omega}, \eqref{ghward},  \eqref{nij}, \eqref{Qij}, \eqref{color} remain valid, up to possible irrelevant linear breaking terms. Consider, for example, the equation of motion of ${\bar \varphi}^a_i$, eq.\eqref{eqmotion-barphi}. With the inclusion  of $\left( \bar{\varphi}_{i}^{a}\varphi_{i}^{a}-\bar{\omega}_{i}^{a}\omega _{i}^{a}\right)$, the breaking term $\Delta^{ai}_{\bar{\varphi}}$, eq.\eqref{delta-barphi}, gets modified by the addition of a linear  term, {\it i.e.}
\begin{equation}
\Delta^{ai}_{\bar{\varphi}} \rightarrow \Delta^{ai}_{\bar{\varphi}} +\mu^2 {\varphi}^a_i \;. \label{blt}
\end{equation}
As such, the Ward identity \eqref{eqmotion-barphi} retains its full validity. A similar argument applies to the other identities. Of the whole set of Ward identities established in the previous sections, only the two Ward identities in eqs.\eqref{int1} and \eqref{int2} cannot be maintained due to the introduction of the term $\left( \bar{\varphi}_{i}^{a}\varphi_{i}^{a}-\bar{\omega}_{i}^{a}\omega _{i}^{a}\right)$. Nevertheless, it turns out that the lack of these two Ward identities does not prevent us to prove the renormalizability of the model. \\\\The second term, $A^{a}_{\mu}A^{a}_{\mu}$, in expression \eqref{RGZ} is more delicate, since it is not BRST invariant
\begin{equation}
s\int d^{4}x\,\frac{1}{2}A^{a}_{\mu }A^{a}_{\mu }= - \int d^{4}x\,A^{a}_{\mu }\partial
_{\mu }c^{a}\,. \label{sbm}
\end{equation}
Moreover, as the induced breaking is a soft breaking, we can repeat the previous argument and convert this breaking into a linear one. Let us briefly sketch how this is done. We introduce a BRST quartet of auxiliary fields $(\mathcal{\tilde{E}}, \mathcal{E}, \sigma, \beta)$, where $(\mathcal{\tilde{E}}, \mathcal{E})$ are a pair of anticommuting fields, while  $( \sigma, \beta)$ are commuting,
\begin{eqnarray}
s\mathcal{\tilde{E}} &=&\sigma\,, \nonumber\\
s\sigma &=&0\,, \nonumber\\
s\beta &=&\mathcal{E}\,, \nonumber\\
s\mathcal{E} &=&0\,.
\end{eqnarray}
Thus, the term $m^2 \int d^4x\; A^{a}_{\mu }A^{a}_{\mu }$ is replaced by the equivalent term $S_m$
\begin{eqnarray}
S_{m} &=&s\int d^{4}x\left( \frac{1}{2}\mathcal{\tilde{E}}\text{ }A_{\mu
}^{a}A_{\mu }^{a}-\mathcal{\tilde{E}}\beta  \right)
+m^{2}\int d^{4}x\,\beta (x) \nonumber\\
&=&\int d^{4}x\left( \frac{1}{2}\sigma A_{\mu }^{a}A_{\mu }^{a}+%
\mathcal{\tilde{E}}A_{\mu }^{a}\partial _{\mu }c^{a}+\mathcal{%
\tilde{E}E}-\beta(\sigma-m^{2})\right) \,.
\end{eqnarray}
Notice in fact that from the equation of motion of $\beta$ we have:
\begin{equation}
\frac{\delta S_{m}}{\delta\beta}=-(\sigma-m^{2})=0\,.
\end{equation}
Therefore,  making the substitution $\sigma = m^2$, it follows
\begin{equation}
S_{m}=\int d^{4}x\left( \frac{m^{2}}{2} A_{\mu }^{a}A_{\mu }^{a}
+\mathcal{\tilde{E}}(A_{\mu }^{a}\,\partial _{\mu }c^{a}+\mathcal{
E})\right)\,.
\end{equation}
Now, performing a linear shift in the $\mathcal{E}$ variable, with unity Jacobian,
\begin{equation}
\mathcal{E}\to\mathcal{E}-A^{a}_{\mu}\,\partial_{\mu}c^{a}\,,
\end{equation}
we obtain
\begin{equation}
S_{m}\to\int d^{4}x\left( \frac{m^{2}}{2} A_{\mu }^{a}A_{\mu }^{a} \right)
+\int d^{4}x\,\mathcal{\tilde{E}}\mathcal{
E}\,.
\end{equation}
As the last term,  $\int d^{4}x\,\mathcal{\tilde{E}}\mathcal{
E}$, is completely decoupled from the rest, the fields $(\mathcal{\tilde{E}},\mathcal{
E})$ are harmless and  can be integrated out, so that the starting term $m^2 \int d^4x\; A^{a}_{\mu }A^{a}_{\mu }$ is recovered. The soft breaking in eq.\eqref{sbm} has thus been converted into a linear breaking, as it is apparent from
\begin{equation}
s{S}_{m}=m^{2}\int d^{4}x\,\mathcal{E}(x)\,. \label{linm}
\end{equation}
The linear breaking \eqref{linm} can be now easily incorporated into the Ward identities \eqref{st1}, \eqref{eqmotion1}, \eqref{eqmotion2}, \eqref{eqmotion3}, \eqref{nonrenorm},  \eqref{eqmotion-barphi}, \eqref{eqmotion-baromega}, \eqref{eqmotion-phi}, \eqref{eqmotion-omega}, \eqref{ghward},  \eqref{nij}, \eqref{Qij}, \eqref{color}, and the previous algebraic analysis can be repeated. We limit here only to state the final result, confirming that the refined action \eqref{RGZ} is renormalizable within the linear breaking formulation. In particular, the nonrenormalization theorem of the ghost-antighost-gluon vertex, $Z_g Z_cZ^{1/2}_A=1$, and the relation $Z_{\gamma^2}= Z^{-1/2}_g Z^{-1/4}_A$ remain valid.

\section{Considerations on the breaking of the BRST symmetry, Gribov horizon and dimension two condensates}

In this section we provide a few remarks on the breaking of the BRST symmetry within the context of the Refined Gribov-Zwanziger framework,  obtained by taking into account the nonperturbative effects of the Gribov horizon and of the dimension two condensates $\langle A^a_\mu A^a_\mu \rangle$, $\langle {\bar \varphi}^{a}_i \varphi^{a}_i -   {\bar \omega}^{a}_i \omega^{a}_i \rangle$.  As we shall see, the Refined Gribov-Zwanziger set up enables us to argue that the breaking of the BRST symmetry might not be in conflict with the confining character of the theory. Rather, it allows us to put forward the idea that the nonperturbative effects of the Gribov horizon and of the dimension two condensates conspire in such a way that the resulting BRST breaking does not spoil at all the consistency of the theory. \\\\Let us start by reminding that, in the Refined Gribov-Zwanziger framework, the gluon propagator, eq.\eqref{rgzgluon},  violates positivity while attaining a nonvanishing value at the origin \cite{Dudal:2007cw,Dudal:2008sp}.  Concerning the ghost propagator, it essentially keeps a free behavior at very low energies  \cite{Dudal:2007cw,Dudal:2008sp}, namely
\begin{equation}
\langle c^a(k) {\bar c}^{b}(-k) \rangle_{RGZ} \Big|_{k\approx  0} \approx \frac{\delta^{ab}}{k^2}  \;. \label{rgzghost}
\end{equation}
As already mentioned,   this behavior of the gluon and ghost propagators is in good agreement with the most recent lattice data \cite{Cucchieri:2007rg,Cucchieri:2008fc,Bornyakov:2009ug,Dudal:2010tf}. In particular, expression \eqref{rgzgluon} provides an accurate fit of the lattice data up to $1.5\;GeV$ \cite{Dudal:2010tf}. To some extent, expression  \eqref{rgzghost} can be taken as evidence of the fact ghosts behave freely at very low energies, {\it i.e.} they decouple in the deep infrared.\\\\This feature is corroborated by the observation that, in the Refined Gribov-Zwanziger framework,  the renormalization group invariant ghost-antighost-gluon effective coupling\footnote{We remind here that the possibility of introducing the RGE invariant quantity  $\alpha_{gh-gl}(k^2) \sim {\cal D}(k^2) {\cal J}^2(k^2)$ relies on the nonperturbative validity of the nonrenormalization theorem of the ghost-antighost-gluon vertex, $Z_c Z_g Z_A^{1/2}=1$, see ref.\cite{Cucchieri:2004sq} for a lattice investigation.}, $\alpha_{gh-gl}(k^2) \sim {\cal D}(k^2) {\cal J}^2(k^2)$, where $\cal D$ and $\cal J$ are the gluon and ghost form factors\footnote{The gluon and ghost form factors, ${\cal D}(k^2), {\cal J}^2(k^2)$, are introduced through the expressions
\begin{equation}
\langle A_{\mu }^{a}(k)A_{\nu }^{b}(-k) \rangle  =  \delta ^{ab}
\left( \delta _{\mu \nu }-\frac{k_{\mu }k_{\nu }}{k^{2}}\right) \frac{{\cal D}(k^2)}{k^2} \;, \qquad
\langle c^a(k) {\bar c}^{b}(-k) \rangle  =   \delta^{ab} \frac{{\cal J}(k^2)}{k^2}  \;. \label{formf}
\end{equation}
} \cite{Cucchieri:2004sq}, goes to zero in the deep infrared, {\it i.e.} $\alpha_{gh-gl}(k^2)\big|_{k \rightarrow 0} \rightarrow 0$.    The coupling $\alpha_{gh-gl}(k^2)$ is directly related to the ghost-antighost-gluon vertex of the Gribov-Zwanziger action and of its refined version. The fact that it goes to zero at very low energies can be seen as a further signal that ghosts get less and less interacting in the deep infrared. It is worth to emphasize that we are not claiming here that the quantity   $\alpha_{gh-gl}(k^2)$, as defined above, has the meaning of the {\it physical coupling constant} of Yang-Mills theory. To our knowledge, the problem of the physical definition of the YM coupling constant is far from being solved and is currently under intensive debate. See  \cite{Aguilar:2010gm}  for a recent proposal. \\\\We are now ready to present our considerations about the breaking of the BRST symmetry within the Refined Gribov-Zwanziger framework.
\subsection{Renormalizability}
The first issue which we point out is that of the renormalizability. One might argue that the breaking of the BRST invariance could jeopardize the renormalizability of the theory. Though, as we have discussed in the previous sections, this is not the case. The breaking of the BRST symmetry induced by the Gribov horizon and by the dimension two condensates is fully compatible with the renormalizability of the theory. This is well captured by the present linear formulation in which the linearly broken BRST invariance can be directly translated into useful Slavnov-Taylor identities, eq.\eqref{st1}, while keeping the nilpotency of the BRST operator $s$, eq.\eqref{nilp}.

\subsection{Gluon confinement}

In a nonconfining gauge theory the characterization of the physical subspace heavily relies on the notion of the BRST charge, $Q_{BRST}$, and thus on the exact BRST invariance of the theory. We remind here that the BRST charge $Q_{BRST}$ enables us to remove from the asymptotic physical subspace the unphysical polarizations of the gauge field, {\it i.e.} the longitudinal and the temporal modes, as well as the negative norm states associated to ghost and antighost states. The resulting subspace, identified by means of  the cohomology of the BRST charge $Q_{BRST}$ in Fock space, contains only positive norm states corresponding to the physical transverse polarization of the gauge field. All this is well understood and applies to nonconfining theories for which the asymptotic fields can be safely introduced, having a direct relationship with the excitations of the physical spectrum. \\\\In a confining theory things are completely different. Gluons are not part of the physical spectrum, due to color confinement. One needs thus a mechanism which enables us to remove them from the physical spectrum.  Here, we argue that such mechanism is provided by the combined effects of the Gribov horizon and of the dimension two condensates. These effects conspire in such a way that gluons become unphysical  in the low energy region, as it follows  from the expression of the gluon propagator \eqref{rgzgluon} which exhibit positivity violation, while displaying  complex poles \cite{Dudal:2007cw,Dudal:2008sp}. As a consequence, transverse gluons cannot correspond to physical excitations of the spectrum. On the other hand, in the Refined Gribov-Zwanziger framework, ghosts seem to decouple  at very low energies. It should be pointed out that the free behavior of the ghosts in the deep infrared is essentially due to the dimension two condensates. Neglecting these condensates, would lead to a very different behavior of the ghosts in the infrared. In fact, the ghost propagator would be enhanced in the deep infrared, $\langle c^a(k) {\bar c}^{b}(-k) \rangle_{GZ}|_{k\approx  0} \approx \frac{\delta^{ab}}{k^4}$, while the ghost-antighost-gluon effective coupling $\alpha_{gh-gl}(k^2)$ would attain a nonvanishing value, {\it i.e.}  $\alpha_{gh-gl}(k^2)\big|_{k \rightarrow 0} \rightarrow \alpha_c \neq 0$ \cite{Alkofer:2000wg}. All this indicates that, without the inclusion of the dimension two condensates, ghosts would be strongly interacting in the infrared. In summary, in the Refined Gribov-Zwanziger framework, the Gribov horizon and the dimension two condensates provide us a  mechanism for the removal of the transverse gluons from the spectrum, while turning ghosts almost free at very low energies. In this sense, we would  have  a tendency to believe that the breaking of the BRST symmetry does not compromise the consistency of the theory.

\subsection{The physical spectrum}
We come now to the pivotal issue to be addressed by any framework aiming at facing a confining theory: the analytic characterization  of the physical spectrum. Even if a possible mechanism for the removal of the gluons from the spectrum of the theory has been worked out, this has to be regarded only as a preliminary step. The real challenge is the physical spectrum of the theory and the analytic calculation of the masses of the physical excitations which, in the present quarkless case, are glueballs \cite{Mathieu:2008me}. \\\\From a theoretical and first principle point of view, the physical spectrum of the theory can be accessed through the evaluation of the correlation functions, $\langle O(k) O(-k) \rangle $, of a suitable set of local composite operators $\{ O \}$. This is a highly nontrivial task, given that the correlator $\langle O(k) O(-k) \rangle $ has to be evaluated with a confining, positivity violating, gluon propagator which exhibit complex poles, see expression \eqref{rgzgluon}.  \\\\On physical grounds, we expect that, in order to be a good candidate, a local operator $O$ should display several features. Of course, we cannot be exhaustive here. We quote, for example, the following properties:
\begin{itemize}
\item $i)$ the operator $O$ should be a colorless operator and we should be able to consistently evaluate the correlation function $\langle O(k) O(-k) \rangle $. We look thus at renormalizable operators.

\item $ii)$ the correlation function should satisfy the K{\"a}ll{\'e}n-Lehmann spectral representation
\begin{equation}
\langle O(k) O(-k) \rangle = \int_0^\infty d\tau\; \frac{\rho(\tau)}{k^2+\tau} \;, \label{spectr}
\end{equation}
with a positive spectral density $\rho(\tau)$.
\end{itemize}
Condition $i)$ is a basic requirement. Certainly, the BRST plays a very important role here. We remind  in fact that, even in the presence of the breaking, the operator $s$ keeps its nilpotency, $s^2=0$, eq.\eqref{nilp}, so that one still has the notion of the BRST cohomology. Moreover, looking at the BRST transformations, eqs.\eqref{BRS}, \eqref{BRS2}, one realizes that all fields $( \bar{c}^{a}, b^a, \bar{\omega}_{\mu }^{ab}, \omega _{\mu }^{ab}, \bar{\varphi}_{\mu }^{ab}, \varphi _{\mu }^{ab} )$
and $(\bar{\mathcal{C}}_{\mu \nu }^{ab},\lambda _{\mu \nu }^{ab},\eta
_{\mu \nu }^{ab},\mathcal{C}_{\mu \nu }^{ab}, \bar{\rho}_{\mu \nu
}^{ab},\bar{\lambda}_{\mu \nu }^{ab},\bar{\eta}_{\mu \nu }^{ab},\rho _{\mu
\nu }^{ab})$ give rise to BRST doublets \cite{Piguet:1995er}. As a consequence, the cohomology classes of the BRST operator with zero ghost number are given by gauge invariant colorless operators built with the field strength $F^a_{\mu\nu}$ and its covariant derivative \cite{Piguet:1995er}. Although  we lack exact BRST invariance, there is an important observation which can be made about the local composite operators belonging to the cohomology of $s$. Remarkably, these operators can be proven to be renormalizable, and this even in the presence of the BRST breaking, see \cite{Dudal:2009zh} for a detailed discussion of the renormalizability of the operator $F^2(x)= F^a_{\mu\nu}(x) F^a_{\mu\nu}(x)$. In other words, the existence of a BRST breaking does not invalidate the renormalizability of the gauge invariant operators belonging to the cohomology of the BRST operator. This feature can be taken as a support in favor of the fact that operators belonging to the BRST cohomology might be useful  for the investigation of the spectrum of the theory  \cite{Sorella:2010it,Capri:2010pg,Dudal:2010cd}. Evidently, due to the presence of the breaking, the BRST exact piece of the renormalized version $O_{ren}$ of a given operator $O$ has to be treated with due care when computing the renormalized correlator $\langle O_{ren}(k) O_{ren}(-k) \rangle $.\\\\Nevertheless, to our understanding, the requirement of renormalizability represents only a first step towards the identification of a meaningful composite operator. The fulfillment of the spectral representation, eq.\eqref{spectr}, with a positive spectral density seems to be a key ingredient in order to provide a link with the physical spectrum. In fact, the knowledge of the spectral density
\begin{eqnarray}
\rho(\tau) & = & {\cal R}\; \delta(\tau - M^2) + \theta(\tau-\tau_0) {\hat \rho}(\tau)  \;, \nonumber \\
\langle O(k) O(-k) \rangle & =&  \frac{{\cal R}}{k^2+M^2} + \int_{\tau_0}^\infty d\tau\; \frac{{\hat \rho}(\tau)}{k^2+\tau} \;, \label{spectr1}
\end{eqnarray}
would give us access to the mass $M$ of the excitation associated to  the composite operator $O$ and to the threshold $\tau_0$ for the production of multiparticles states.. In addition, the spectral representation \eqref{spectr} will ensure that the correlation function $\langle O(k) O(-k) \rangle $ can be Wick rotated from the Euclidean to the Minkowski space, a feature of fundamental importance for a correct physical interpretation of the theory. Though, till now, the existence of the spectral representation seems  to be uncorrelated with the issue of the BRST symmetry. In particular, it is not easy to figure out a possible relationship between BRST and the positivity of the spectral density. We remind here the fact that the correlation function $\langle O(k) O(-k) \rangle $ has to be evaluated with a confining gluon propagator which violates positivity. From this, one can really appreciate the hard task of accessing the spectrum of the glueballs. In this context, it is worth emphasizing that the analytic structure emerging from the use of the refined gluon propagator \eqref{rgzgluon} looks quite useful. As discussed in \cite{Baulieu:2009ha}, the complex conjugate poles associated to expression  \eqref{rgzgluon} combine in such a way that the final result does display a nice spectral representation with positive spectral densities, at least for a certain class of composite operators. This feature has enabled us to extract numerical estimates for the masses of the three lightest glueballs, $0^{++}, 2^{++}, 0^{-+}$ \cite{Dudal:2010cd}, which turn out to be in good agreement with the available lattice data, see \cite{Mathieu:2008me}. This is an encouraging result, confirming that the interplay between the Gribov horizon and the dimension two condensates might be helpful in order to access the glueball spectrum. Though, we have the impression that, in a confining theory, BRST alone would not be sufficient for a complete characterization of the spectrum. Some additional requirements gauranteeing the validity of the spectral representation and of the positivity of the corresponding spectral density seem to be needed, especially in view of the fact that explicit calculations have to be carried out with a gluon propagator which violates positivity.

\section{Conclusion}

In this work, the soft breaking of the BRST symmetry due to the Gribov horizon and to the dimension two condensates has been investigated. This breaking can be converted into a linear breaking, a feature which has enabled us to write down a useful set of Ward identities. These Ward identities ensure the multiplicative renormalizability of both Gribov-Zwanziger action and its refined version. \\\\We underline that the behavior of the gluon and ghost propagators obtained from the Refined Gribov-Zwanziger framework turns out to be in good agreement with the most recent lattice data. Recently, the refined gluon propagator has been employed to estimate the masses of the three lightest glueballs, $0^{++}, 2^{++}, 0^{-+}$. Also here, the results are in agreement with the available data. \\\\Although much work is certainly needed in order to achieve a satisfactory understanding of the role played by the BRST symmetry in a confining theory, the encouraging results obtained so far within the Refined Gribov-Zwanziger framework have enabled us to put forward a few observations in favor of the idea that the breaking of the BRST symmetry, in both soft and linear formulations, does not jeopardize  the consistency of the theory.

%%%%%%%%%%%%%%%%%%%%%%%%%%
\section*{Acknowledgments}
%%%%%%%%%%%%%%%%%%%%%%%%%%%
The Conselho Nacional de Desenvolvimento Cient\'{\i}fico e Tecnol\'{o}gico
(CNPq-Brazil), the Faperj, Funda{\c{c}}{\~{a}}o de Amparo {\`{a}} Pesquisa
do Estado do Rio de Janeiro, the Latin American Center for Physics (CLAF),
the SR2-UERJ, the Coordena{\c{c}}{\~{a}}o de Aperfei{\c{c}}oamento de
Pessoal de N{\'{\i}}vel Superior (CAPES) are gratefully acknowledged.

\end{document}